\documentstyle[aps,multicol,pre,epsfig]{revtex} 

\newcommand{\beq}{\begin{equation}}  
\newcommand{\eeq}{\end{equation}}   
 
\begin{document}  
\begin{title}
{\bf Avoiding Infrared Catastrophes in Trapped 
Bose-Einstein Condensates} 
\end{title} 
 
\author{P.G. Kevrekidis$^{1}$, G. Theocharis$^2$, 
D.J. Frantzeskakis$^{2}$, and A. Trombettoni$^3$}  
\address{$^{1}$ Department of Mathematics and Statistics, 
University of Massachusetts, Amherst MA 01003-4515, USA \\  
$^{2}$ Department of Physics, University of Athens, Panepistimiopolis,  
Zografos, Athens 15784, Greece \\  
$^3$ Istituto Nazionale per la Fisica della Materia and 
Dipartimento di Fisica, Universita' di Parma, 
parco Area delle Scienze 7A, I-43100 Parma, Italy} 
\maketitle   
 
\begin{abstract}  
This paper is concerned with the long wavelength instabilities 
(infrared catastrophes) occurring in Bose-Einstein condensates (BECs). 
We examine the modulational instability in 
``cigar-shaped'' (1D) attractive BECs and the  
transverse instability of dark solitons in ``pancake'' (2D)  
repulsive BECs. We suggest mechanisms, and give explicit estimates, 
on how to  ``engineer'' the trapping conditions of the condensate to avoid  
such instabilities: the main result being that a tight enough 
trapping potential suppresses the instabilities present  
in the homogeneous limit. We compare the obtained estimates  
with numerical results and we  
highlight the relevant regimes of dynamical behavior. 
\end{abstract}   
 
 
\vspace{2mm}   
 
\section{Introduction and Setup}
 
Infrared catastrophes, or long-wavelength instabilities (LWI) as 
they are otherwise known, are ubiquitous in physical phenomena.  
From Magneto-Hydro-Dynamics  \cite{MHD} to chemical models  
\cite{RD}, and from quantum systems \cite{wire}  
to fluid mechanics \cite{fluids}, polymer  
physics \cite{polymers} and plasmas \cite{plasma}, large  
scale modulations may destabilize the system of interest. 
On the other hand,  
many of these instabilities (and their thresholds) have been  
quantified in the context of nonlinear models. 
 
In the past few years, another context that can be accurately  
modeled by nonlinear partial differential  
equations that are known to possess LWI, has become experimentally tractable. 
This setting is, in particular, the one  
of Bose-Einstein condensates (BECs) \cite{review}  
whose experimental realization has led to an explosion of  
interest in the field of atomic matter waves and of nonlinear excitations  
in them. Such nonlinear waves have been recently experimentally  
generated in BECs, namely dark \cite{dark}  
and bright \cite {bright,paris} solitons. Also,  
two dimensional excitations, such as vortices \cite{vortex} and  
lattice patterns thereof \cite{vl} have also been obtained   
experimentally, while other nonlinear waves,  
such as Faraday waves \cite{stal}, ring   
solitons and vortex necklaces \cite{theo} have been theoretically predicted.  
It is worth noting here that the nature of Bose-Einstein nonlinear matter  
waves depends crucially on the type of the interatomic interactions:  
dark (bright) solitons can be created in BECs with repulsive  
(attractive)  
interatomic interactions, resulting from positive (negative)  
scattering length. 
 
In the present work, our scope is to re-examine the LWI  
in the context of Bose-Einstein condensates and  
exploit the consequences on the LWI of the  
inhomogeneity induced by the trapping potential.  
It is well-known \cite{review} in this setting that the  
condensates are formed under appropriate confining conditions 
that typically consist of magnetic traps modeled by  
parabolic potentials. Our scope is  
then two-fold. On the one hand, we aim at illustrating the potential for  
the instabilities and at examining their dynamical development when  
they exist. On the other hand, the magnetic confinement provides an  
additional ``trapping'' length scale to the problem whose  
{\it competition} with the instability length scale may disallow  
the dynamical manifestation of the LWI. In this way, we will propose 
how to {\it engineer} trapping conditions so as to prevent  
unstable dynamical evolution. 
 
We will examine two benchmark examples of long-wavelength  
instabilities in BECs: the modulational instability  
 occurring in attractive, one-dimensional (1D) BECs and  
giving rise to the formation of bright matter-wave solitons, and  
the transverse (``snaking'') instability of dark soliton stripes  
formed in repulsive two-dimensional (2D) BECs. It should be noted  
here that the genuinely three-dimensional (3D) condensate can be  
considered as approximately 1D if the nonlinear inter-atomic interaction  
is weak relative to the trapping potential force in the transverse  
directions; then, the transverse size of the condensates is much smaller than  
their length, i.e., the BEC is ``cigar-shaped'' and can  
be effectively described by 1D models \cite{GPE1d,VVK}. Similarly, if 
the transverse confinement is strong along one direction and weak 
along the others, then for this ``pancake-shaped'' BEC, 2D model equations 
are relevant \cite{GPE2d}. 
 
Close to zero temperature, it is well-known that the 3D Gross-Pitaevskii (GP)  
equation \cite{review} accurately captures the dynamics of the condensate. 
For cigar-shaped BECs, the model equation is effectively 1D  
and can be expressed in the following dimensionless form, 
\begin{eqnarray} 
i \frac{\partial u}{\partial t}= -  
\frac{1}{2} \frac{\partial^2 u}{\partial x^2} + \alpha |u|^2 u + V(x) u, 
\label{eqn1} 
\end{eqnarray} 
where $u$ is the macroscopic wave function  
(normalized to $1$), and $\alpha=a/|a|=\pm 1$ is the renormalized  
scattering length,  
which is positive (negative) for repulsive (attractive) condensates.  
In this equation, $t$ and $x$ are respectively measured in units of  
$9/16\epsilon^{2}\omega_{\perp}$ and  
$3 a_{\perp}/4 \epsilon$,  
where $\omega_{\perp}$ is the confining frequency in the transverse  
direction, $a_{\perp}=\sqrt{\hbar/m \omega_{\perp}}$ is the transverse  
harmonic-oscillator length and $\epsilon \equiv N |a|/a_{\perp}$ is a  
small dimensionless parameter, $N$ being the total number of atoms  
\cite{VVK}. Finally,   
\begin{equation}
V(x)=\frac{\Omega ^{2}}{2} x^{2}
\label{V}
\end{equation} 
where $\Omega=(9/16\epsilon^{2}) \cdot  
(\omega_x/\omega_{\perp})$ is the frequency of the magnetic trap 
potential $V$ in our dimensionless units, and $\omega_x$ is the axial  
confining frequency. Similarly the 2D model for the pancake-shaped  
condensate assumes the form: 
\begin{eqnarray} 
i \frac{\partial u}{\partial t}= -  
\frac{1}{2} \Delta u + \alpha |u|^2 u + V(r) u 
\label{eqn2} 
\end{eqnarray} 
where now the role of the axial frequency/spatial variable  
is played by the radial one ($r\equiv \sqrt{x^{2}+y^{2}}$)  
and the normalized variables are connected to the dimensional  
ones similarly to the 1D case, but with the role of $\omega_{x}$  
($\omega_{\perp}$) now played by $\omega_{\perp}$ ($\omega_{z}$). 

We structure our presentation as follows: 
in Section II we study the suppression 
of the modulational instability for 1D cigar-shaped BECs 
in presence of a tight confining magnetic potential, while in 
Section III we investigate the transverse instability for 2D trapped BECs. 
In both cases we provide a criterion (in terms of the trap parameters) 
giving the conditions under which LWI can be avoided. 
Section IV is devoted to our concluding remarks.

\section{Modulational Instability}
   
It is well-known (see, e.g., the recent work \cite{zoi}  
and references therein) that, in the absence of external potential,  
the continuous-wave (cw) solution $u=u_{0}\exp(-i\alpha u_{0}^{2}t)$  
of amplitude $u_0$, of Eq. (\ref{eqn1}) becomes modulationally unstable 
when perturbations of wavenumber $k < k_{cr} \equiv 2 |\alpha|^{1/2} u_0$ 
are imposed.  
This can be equivalently interpreted as follows: when length scales 
\begin{eqnarray} 
\lambda > \lambda_{cr} \equiv \frac{\pi}{u_0 \sqrt{|\alpha|}} 
\label{eqn3} 
\end{eqnarray} 
become ``available'' to the system, then the modulation over these scales 
leads the solution to instability.  
 
However, in the presence of the magnetic trap, 
there is a characteristic scale 
set by the trap, namely the BEC axial size, $\lambda_{BEC}$,  
which depends on the trapping frequency $\Omega$. When  
$\lambda_{BEC} < \lambda_{cr}$ suppression of the modulational  
instability is expected. To estimate $\lambda_{BEC}$ in a specific setup,  
we will examine a protocol relevant to the recent experiments conducted  
by the Rice \cite{bright} and Paris \cite{paris} groups, which has stimulated a considerable amount of theoretical attention  
\cite{brand}. In particular, we start with a 1D repulsive condensate,  
whose ground state wavefunction  
is approximately in the so-called Thomas-Fermi (TF)  
regime \cite{review}, and subsequently change the interaction  
into an attractive one.  
This is experimentally achieved using the so-called Feshbach resonance:  
an external magnetic field is used to modify the scattering length of the  
interatomic interactions \cite{inouye}.  
In our example, we use as initial condition  
the ground state of Eq. (\ref{eqn1}) with $\alpha=1$,  
which is in the TF approximation $|u|^2 \approx \mu - V(x)$;  
then at $t=0$ we change the sign of the scattering length,  
i.e., we set $\alpha=-1$.  
Therefore in this situation $\lambda_{BEC} \approx 2\sqrt{2\mu}/\Omega$. From  
the normalization condition $\int dx |u|^2=1$  
one gets 
\begin{equation}
\mu=\Bigg(\frac{3 \Omega}{4\sqrt{2}} \Bigg)^{2/3}.
\label{mu_1D}
\end{equation}
  
The condition for the suppression of the modulational instability 
$\lambda_{BEC} < \lambda_{cr}$ [where $\lambda_{cr}$ is given by 
Eq. (\ref{eqn3})] gives $\Omega> 2^{3/2} (|\alpha| \mu)^{1/2} u_0/\pi$. 
In this context, the amplitude $u_{0}$ in Eq. (\ref{eqn5})  
can be well approximated as $u_0 \approx \sqrt{\mu}$,  
since the TF approximation is most accurate close to the center  
of the condensate. Therefore if $\Omega > \Omega_{cr}$, where 
\begin{eqnarray} 
\Omega_{cr} = \frac{9}{\sqrt{2} \pi^3} \approx 0.2, 
\label{eqn5} 
\end{eqnarray} 
then the trapping conditions are ``engineered'' in such a way that  
the modulational instability cannot manifest itself: 
for a tight enough trapping potential the modulational instability 
does not occur. This is one of the key aims of this paper, 
namely to quantitatively highlight how the infrared catastrophes 
can be avoided in the presence of  
sufficiently ``tight'' trapping of the condensate. We remark 
that if one evaluates $\lambda_{BEC}$ by using for the ground state  
of Eq. (\ref{eqn1}) (with $\alpha=1$) 
a gaussian with variationally determined  
width, one has an estimate of $\Omega_{cr}$ in good agreement  
with Eq. (\ref{eqn5}). We also notice that the critical length 
$\lambda_{cr}$ is a few healing lengths $\xi$: indeed, when the 
the density grows from $0$ to $u_{0}^{2}$ within a distance $\xi$, 
the quantum pressure and interaction energy terms are equal when
$1/(2\xi^{2}) \approx u_{0}^{2}$, which means that
$\xi \approx (\sqrt{2}u_{0})^{-1}$ and therefore 
$\lambda_{cr} \approx \pi\sqrt{2}\xi = 4.44 \xi$. 

Typical experimental values for a $^{7}$Li BEC are  
$\omega_{x}=2\pi \times 5$Hz, $\omega_{\perp}=2\pi \times 500$Hz,  
$a=-3a_0$ and $N \approx 10^{3}$; these yield  
$\epsilon \sim 0.1$ and $\Omega \sim 0.5$, which should be  
a sufficiently 
large value to avoid the emergence of the modulational instability. 
 
\begin{figure} 
\centerline{\psfig{figure=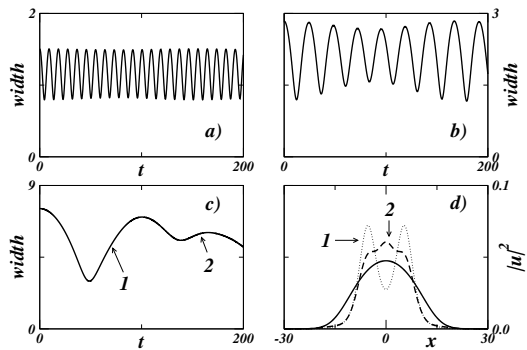,width=60mm,angle=270}} 
\caption{Condensate width as a function of time for  
$\Omega=0.3$ (a), $\Omega=0.1$ (b), and $\Omega=0.02$ (c).  
In (d) we plot the atomic density with $\Omega=0.02$  
at three different times:  
$t=0$ (solid line), $t=80$ (dotted line, corresponding  
to point $1$ in (c)) and $t=160$ (dashed line, corresponding  
to point $2$ in (c)).} 
\label{fig1} 
\end{figure}

We tested this prediction by means of direct numerical simulations  
shown in Fig. \ref{fig1}. The panels (a), (b), (c) show respectively  
the case of $\Omega=0.3$, $0.1$, and $0.02$. 
It is seen that the supercritical, ``tight'' trap 
$\Omega=0.3$ does not allow the development of the instability 
for the condensate [Fig. \ref{fig1} (a)] .  
The only consequence of the change of the sign of the scattering length is  
the excitation of an internal mode of the condensate, due to the fact  
that the TF cloud is not the ground-state  
for $\alpha=-1$. This results in nearly periodic oscillations of  
the width of the wavefunction  
$W=[\int x^2 |u|^2 dx - (\int x |u|^2)]^{1/2}$,  
which we monitor as a diagnostic in our numerical examples.  
For $\Omega=0.1$ [Fig. \ref{fig1} (b)], the width oscillations  
are no longer periodic: the result of the 
modulational instability can be clearly discerned in the addition of  
an extra frequency to the motion of the condensate for subcritical trapping.  
We have identified this to be a rather smooth transition,  
becoming more pronounced close to the theoretically predicted  
critical point of $\Omega \approx 0.2$.  
We have also examined the development of the 
instability for weaker trapping conditions  
[$\Omega=0.02$, see Figs. \ref{fig1} (c), (d)].  
Then an additional frequency emerges,  
as the condensate is now cleaved in two pieces during  
the time evolution [see Fig. \ref{fig1} (d)].  
For even weaker trappings, we found  
that the larger the ratio of $\lambda_{BEC}/\lambda_{cr}$,  
the more frequencies appear in the condensate density evolution,  
rendering it increasingly chaotic. Hence, we have demonstrated that a  
cascade of frequencies arises in the BEC dynamics  due to the modulational  
instability and subsequent higher resonances that allow the breakup of the 
condensate and make its motion less regular.  
This mechanism should clearly be experimentally detectable;  
in fact we surmise that the observation of a single matter-wave 
bright soliton \cite{paris}, rather than a soliton train \cite{bright}, 
is due to the fact that $\lambda_{BEC}$ was smaller in the former case, 
as the number of atoms was an order of magnitude 
fewer in the Paris experiment, while the ratio of the trapping 
frequencies was approximately the same.  
 
 
\section{Transverse Instability}
  
One of the infrared catastrophes that occur in two spatial  
dimensions is the transverse instability of  
dark-soliton stripes in repulsive ($\alpha > 0$) BECs.  
As a result, a dark-soliton undergoes a  
transverse snake deformation \cite{luther},  
causing the nodal plane to decay into vortex pairs.  
This instability has been examined in the context of BECs  
in \cite{shlyap}, and the Bogoliubov spectrum of the dark soliton  
has been obtained; in this context, the relevant imaginary modes  
were identified to transfer the energy of the condensate to collective  
excitations parallel to the nodal plane destroying the configuration.  
However, as is highlighted in \cite{feder}, ``the explicit connection  
between the existence of imaginary excitations and a dynamical snake  
instability remains unclear''. Our scope here is to illustrate  
the criterion for the transverse instability and to 
test it against direct numerical simulations, exposing the possible  
dynamical scenarios and quantifying their dependence 
on the trapping parameters. 
 
In the absence of the potential, the transverse instability occurs  
for perturbation wavenumbers   
\begin{equation}
k < k_{cr} \equiv  
\left[ 2 \sqrt{\sin^2{\phi}+ u_{0}^{-2}\sin{\phi}+
u_{0}^{-4}}-\left(2\sin{\phi}+u_{0}^{-2} \right)\right]^{1/2},
\label{k_cr_2D}
\end{equation}  
where $\sin\phi$ is the dark-soliton velocity 
\cite{luther} and $u_{0}$ is the amplitude of the homogeneous 
background, connected with the chemical potential through 
$u_{0}^{2}=\mu$ similarly to the previous ($1$D) setting.  
In the case of stationary (black) solitons, of interest here,  
$\sin\phi=0$, hence $k_{cr}=u_{0}^{-1}$. On the other hand, for  
$V(r)=\Omega^2 r^2/2$, a similar calculation as for the  
1D problem yields the characteristic length scale of the BEC  
(i.e., the diameter of the TF cloud) as  
$\lambda_{BEC} \approx 2 \sqrt{2\mu}/\Omega$.  
Then, the criterion for the suppression of the transverse instability  
is that the scale of the BEC is shorter than the minimal one  
for the instability. The corresponding condition reads 
\begin{eqnarray} 
\Omega > \frac{\sqrt{2\mu}}{\pi u_{0}}. 
\label{eqn6} 
\end{eqnarray} 
To obtain the minimum value of $\Omega$ we need to know how 
$\mu$ is connected with $u_{0}$. As a first guess, 
in the absence of the dark soliton, 
one can assume $u_{0}^{2} \approx \mu$ 
(close to the center of the BEC), which yields $\Omega > \sqrt{2}/\pi = 
0.45$. Hence, stronger trapping should ``drown'' the transverse instability 
and preserve dark soliton stripes on top of the Thomas-Fermi cloud  
(i.e., stable ``dipole'' solutions). Note that in terms of real  
physical units, the above mentioned critical value of $\Omega$  
may correspond, e.g., to a weakly interacting $^{87}$Rb pancake  
condensate, containing $\approx 10^{3}$ atoms, confined in a trap  
with $\omega_{r}=2\pi \times 5$Hz and $\omega_{z}=2\pi \times 50$Hz. 
 
\begin{figure}[tbp]  
\epsfxsize=6.5cm  
\epsffile{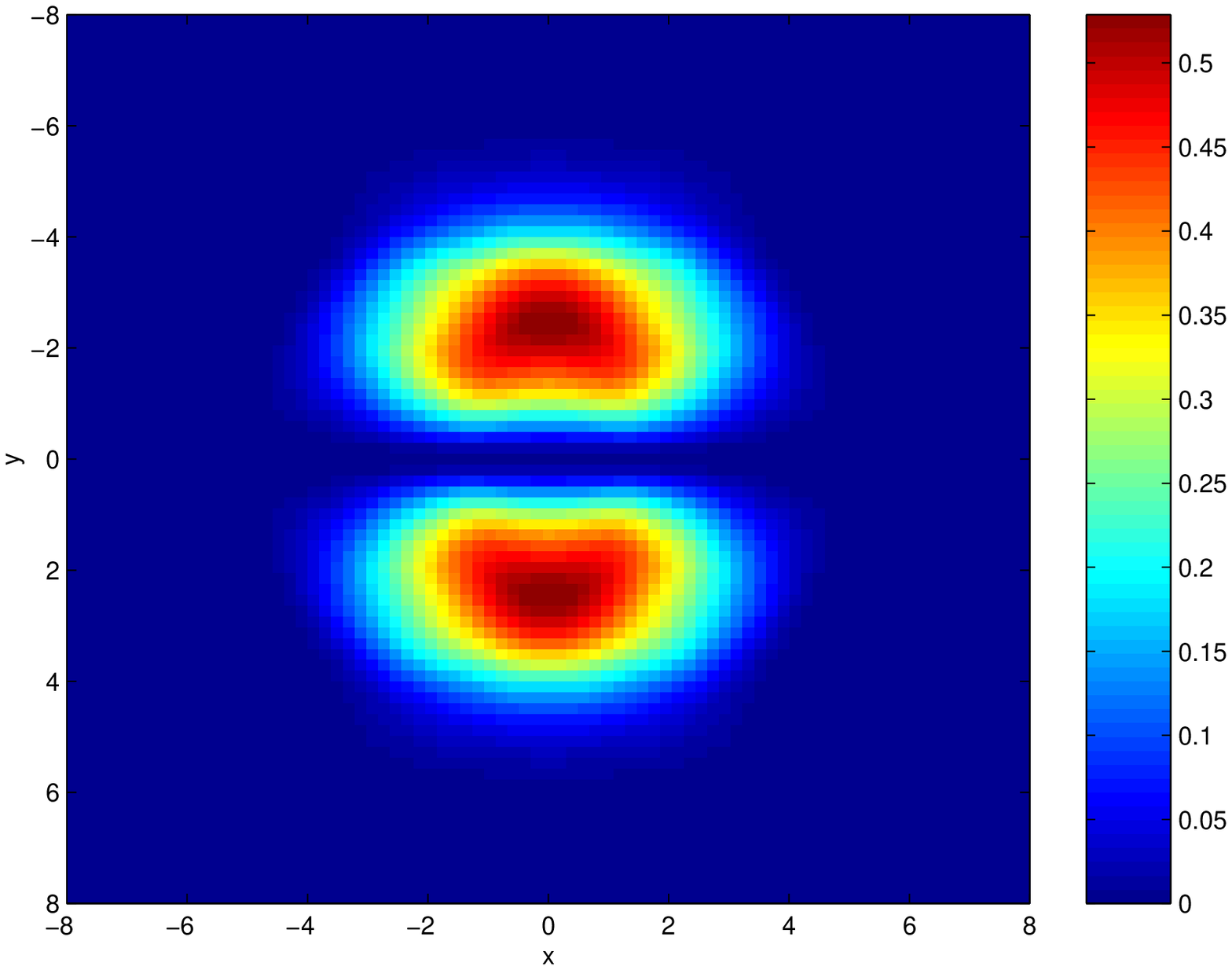} 
\epsfxsize=6.5cm  
\epsffile{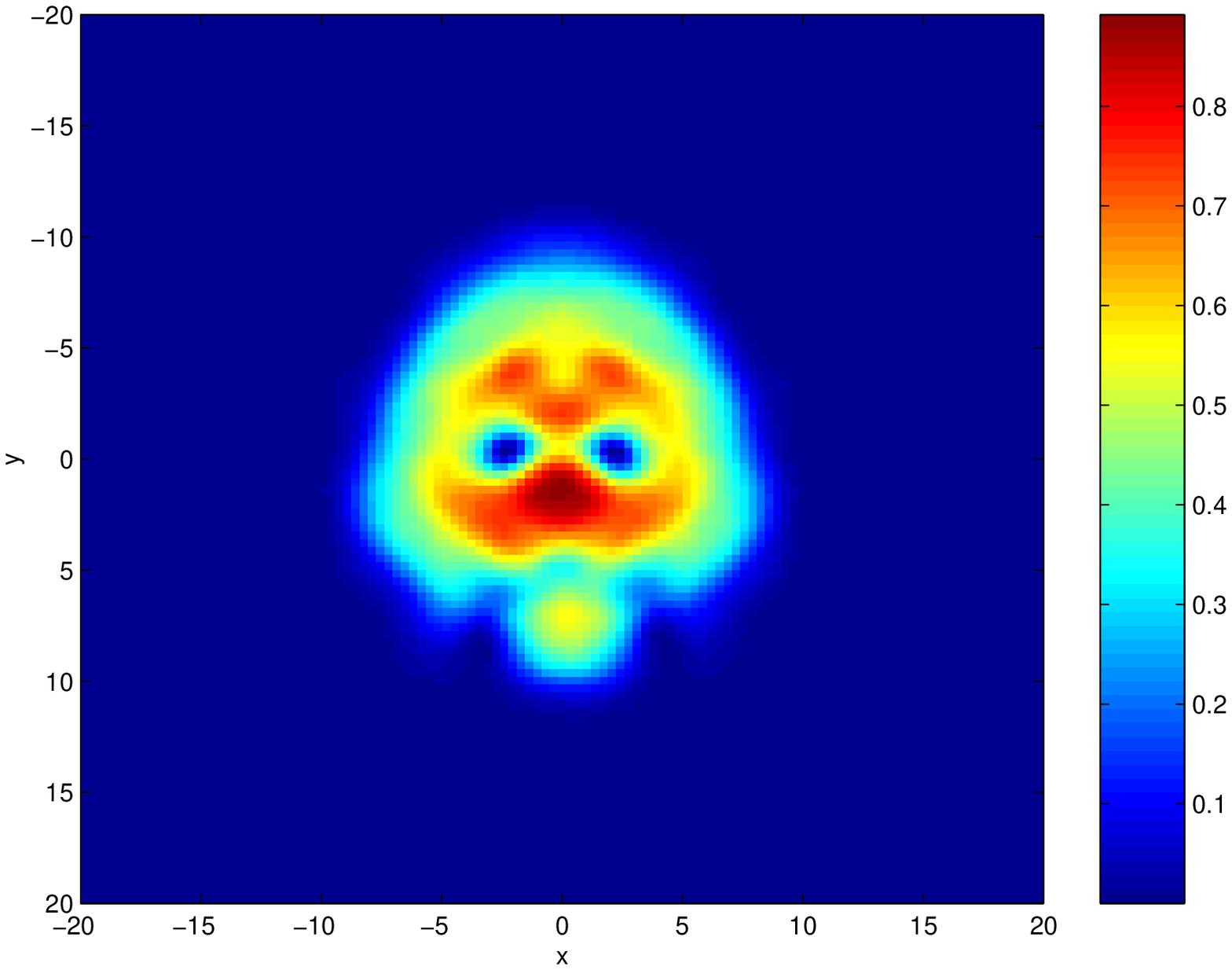} 
\caption{The panels show the contour plots of the density  
$|u|^2$ for $\Omega=0.35$ at $t=1000$ (left) and  
$\Omega=0.15$ at $t=200$ (right). In the first case the  
transverse instability is clearly suppressed, while  
in the second it sets in, giving rise to a formation of a vortex pair. }  
\label{fig2} 
\end{figure}

\begin{figure}[tbp]  
\epsfxsize=6.5cm  
\epsffile{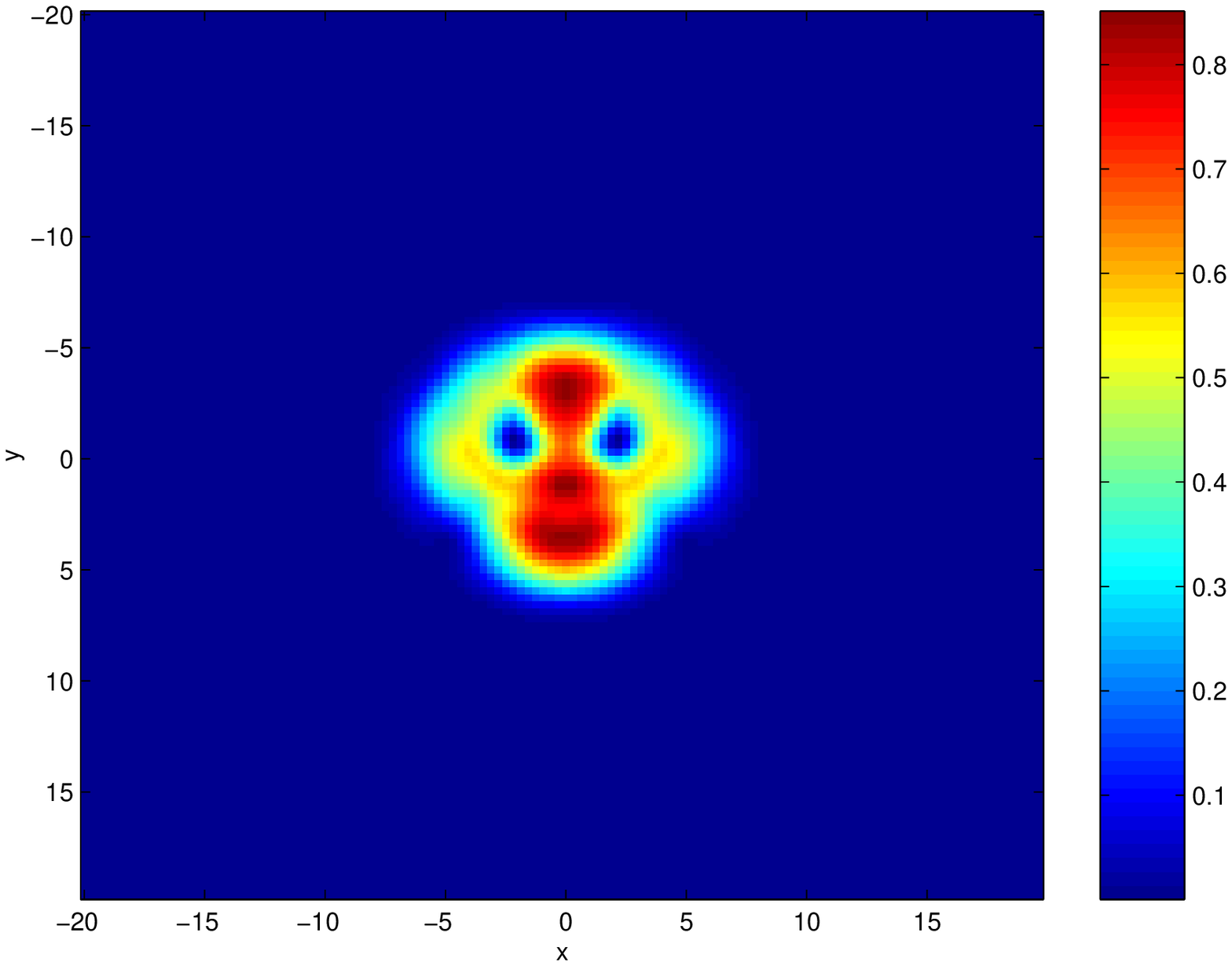} 
\epsfxsize=6.5cm  
\epsffile{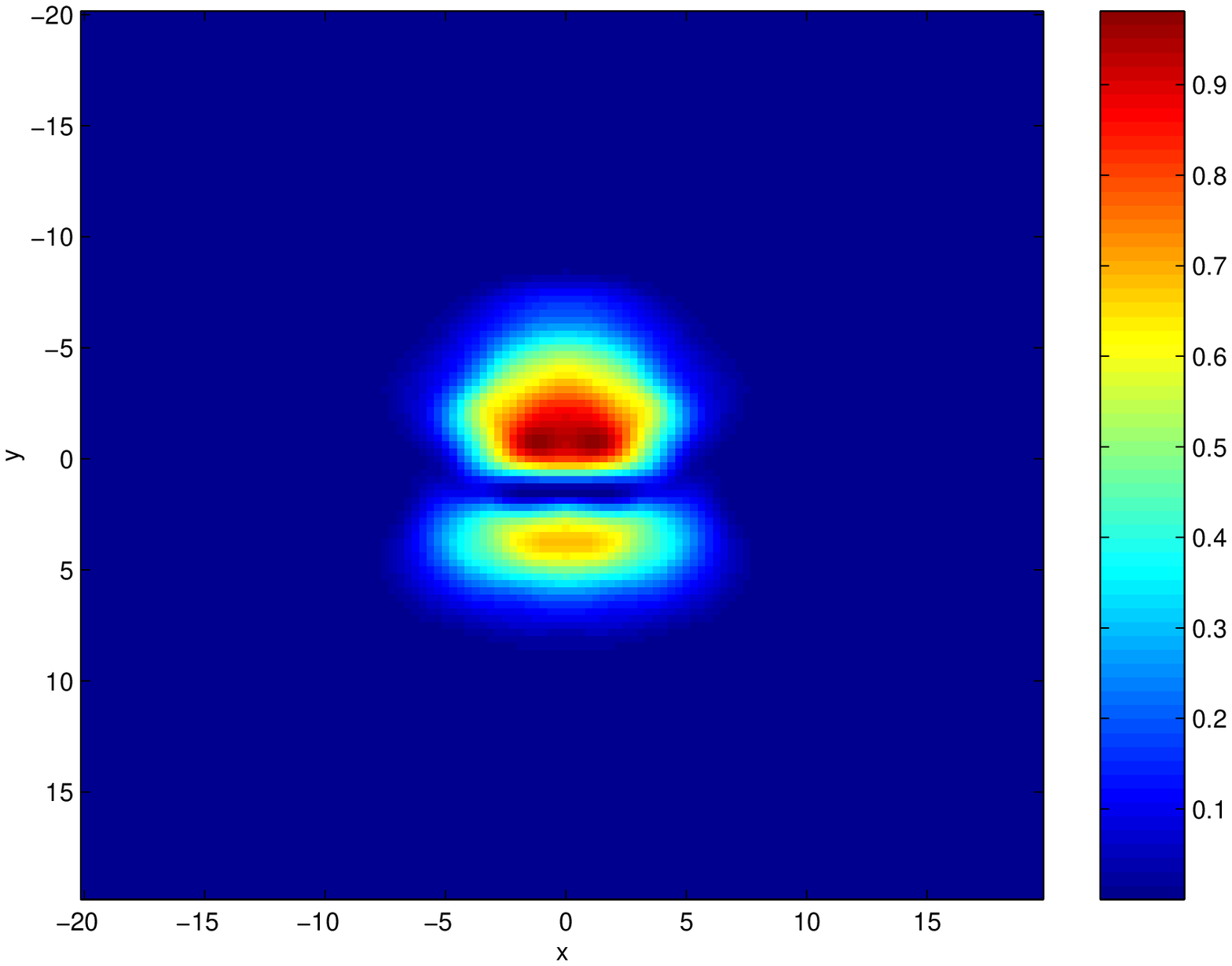} 
\caption{Snapshots of a vortex-pair evolution in a case  
where snaking instability has set in ($\Omega=0.2$).  
In the left panel ($t=190$) the formed vortex pair is shown,  
while the right panel ($t=210$) shows the recombination  
of the two vortices, resulting in the re-generation of a dark stripe 
structure.  
The latter is unstable and decays at longer times ($t\approx 400$). }  
\label{fig3} 
\end{figure}

We have numerically tested this condition, finding it to be an  
{\it overestimate} of the critical trapping frequency  
for the transverse instability, which is $\Omega_{cr}  
\approx 0.31$.     
This result shows that the numerically found  
Fig. \ref{fig2} shows a dynamical evolution example  
of the dipolar solution, initialized with a $\tanh(y)$ imposed on the  
TF cloud, for $\Omega=0.35$ (left) and $\Omega=0.15$ (right).  
Both snapshots show the contour plot of the square modulus of the wave  
function at $t=1000$ (left) and $t=200$ (right).  
Clearly, in the former case the transverse instability is suppressed,  
while in the latter a vortex pair has been formed demonstrating  
the dynamical instability of the configuration.  
We have monitored the asymptotic, long time evolution of the  
dipole and have observed the following interesting phenomenology:  
for $0.18 \lesssim \Omega \lesssim 0.31$, while the stripe is dynamically  
unstable, there is not sufficient space  
for the instability-induced vortices to fully develop; as a result,  
after their formation, they subsequently recombine and disappear.  
This behavior is shown in Fig. \ref{fig3} (for $\Omega=0.2$):  
the two vortices formed are shown at $t=190$ (left panel),  
while at $t=210$ they recombine to form a transient dark stripe  
(right panel). This configuration is unstable and it subsequently  
breaks up to a new vortex pair (not shown here), which eventually  
recombines at longer times ($t\approx 400$). It is interesting  
to note that as, in the present case, the available size of the  
condensate is of the order of a few healing lengths $\xi$  
(in our units, $\xi=1/\sqrt{2}$), the two vortices formed  
hardly fit the condensate size (recall that the vortex core  
is of order of $\xi$ \cite{fetter}). This is a possible  
qualitative explanation of this recombination, whose origin is 
the competition of the length scales available in the BEC.  
Finally, it should be noted that for $\Omega \lesssim 0.18$, 
the vortices will 
survive in the asymptotic evolution of the system, and naturally the weaker  
the trapping the larger the number of ``engulfed'' vortices  
generated due to the stripe breakup.  
 
A question that naturally arises in the results 
above concerns the disparity between the critical 
point theoretical estimate for the transverse instability 
and the corresponding numerical finding. 
We believe that the disparity is justified by the fact  
that the theoretical stability analysis of \cite{luther} is  
performed for the infinite homogeneous medium (the dark-soliton pedestal),  
in the absence of a magnetic trap. In the presence of the trap   
on the one hand, the background is inhomogeneous, 
while, on the other hand,
for tight traps resulting in small condensate sizes, 
the presence of the dark soliton at the BEC center, 
significantly modifies the maximum density. 
Thus, one should expect that the relation $u_{0}^{2}=\mu$ 
should be modified as $\delta u_{0}^{2}=\mu$, where 
the ``rescaling'' factor $\delta<1$. Based on the results 
of the numerical simulations (see Figs. \ref{fig2}-\ref{fig3}) 
we have observed that embedding the dark soliton in the BEC 
center reduces the maximum density of the TF cloud to 
the half of its initial value. 
This suggests that $\delta=1/2$, which, according to Eq. 
(\ref{eqn6}), leads to a new value for $\Omega$ minimum, 
i.e., $\Omega=1/\pi\approx 0.318$. 
This modified criterion for the suppression of the 
transverse instability, namely $\Omega>0.318$ is in 
very good agreement with the above mentioned numerically 
found condition $\Omega>0.31$. 
 
\section{Conclusions}

We have examined case examples of long  
wavelength instabilities in Bose-Einstein condensates.  
We have revisited the modulational instability  
in quasi-1D, cigar-shaped BECs, and the transverse instability 
in quasi-2D, pancake-shaped BECs. We have advocated that trapping  
conditions can be  
engineered to avoid or induce such instabilities at will.  
Using the length scale competition of the entrainment due to trapping and of  
critical instability wavelength, we have given explicit estimates for  
critical values of the trapping frequency beyond which the instabilities  
will be absent.  
We have tested these criteria  
in both cases and have found good agreement with the numerical results  
in the 1D case. 
In the 2D setting, we have explained the overestimation of the  
critical point, on the basis of the  
homogeneous background assumed in the theoretical estimate, 
as well as the modification of the value of the maximum density 
of the condensate due to the presence of the dark soliton. 
Our results demonstrate how to engineer 
the trapping conditions, in order to achieve supercritical regimes, 
devoid of long  
wavelength instabilities. On the other hand, for the subcritical regimes,  
we have illustrated the relevant phenomenology through direct numerical  
simulations and the cascade that leads to 
the instability and eventual destruction of the original coherent  
structure. Such results can be particularly useful in quantifying the  
selection of external  
conditions so as to achieve or avoid a given  experimental outcome.  
 
This work was supported by the Eppley Foundation and 
NSF-DMS-0204585 (PGK), the ``A.S. Onasis'' 
Public Benefit Foundation (GT) and the Special Research Account of the  
University of Athens (GT, DJF), and by M.I.U.R. through grant No. 
2001028294 (AT).

 
 
\end{document}